# National-scale research performance assessment at the individual level[1]

*Giovanni Abramo*[a,b,*], *Ciriaco Andrea D'Angelo*[a]

[a] Laboratory for Studies of Research and Technology Transfer
School of Engineering, Department of Management
University of Rome "Tor Vergata"

[b] National Research Council of Italy

**Abstract**

There is an evident and rapid trend towards the adoption of evaluation exercises for national research systems for purposes, among others, of improving allocative efficiency in public funding of individual institutions. However the desired macroeconomic aims could be compromised if internal redistribution of government resources within each research institution does not follow a consistent logic: the intended effects of national evaluation systems can result only if a "funds for quality" rule is followed at all levels of decision-making. The objective of this study is to propose a bibliometric methodology for: i) large-scale comparative evaluation of research performance by individual scientists, research groups and departments within research institution, to inform selective funding allocations; and ii) assessment of strengths and weaknesses by field of research, to inform strategic planning and control. The proposed methodology has been applied to the hard science disciplines of the Italian university research system for the period 2004-2006.

* **Corresponding author**, Dipartimento di Ingegneria dell'Impresa, Università degli Studi di Roma "Tor Vergata", Via del Politecnico 1, 00133 Rome - ITALY, tel. +39 06 72597362, abramo@disp.uniroma2.it

# 1. Introduction

Over the last two decades, many industrialized countries have introduced national exercises for the evaluation of research activity, responding to demands for greater accountability and for improved allocative efficiency in funding for institutions. Governments and their national agencies are gradually imposing elements of competition in the allocation of public funds. Examples are seen in national systems of resource allocation based on evaluations of project proposals, and also in implementation of systems of "formula funding" based on comparative performance measures. The United States offers an example of the first case: here, financing for research is awarded on a competitive basis, primarily for projects. Meanwhile, the most significant experience of the case of comparative performance measures has been the Research Assessment Exercise in Great Britain, where the fifth edition of the exercise has been concluded (RAE, 2008). The aim is to assess the quality profiles of all UK higher education institutions and use them in allocating not less than 25% of the total government funding for universities, with effect from 2009-10. Similar exercises are also used in other English-speaking nations, most prominently: Excellence in Research for Australia Initiative (ERA) and New Zealand's Performance-Based Research Fund (PBRF, 2008). In Italy, the first Triennial Research Evaluation (VTR, 2006) was carried out in 2006 and the next one (the "VQR") is expected shortly. Here, the intention of the Italian government is to allocate a growing portion of its university research funding (30% in 2011) on the basis of results from the national evaluation.

The various national funding agencies involved have made continuous efforts to improve the methods for their assessments. Until recently they had usually adopted peer review approaches, but lately there has been a tendency towards adoption of quantitative proxies, with the inclusion of bibliometric indicators, where these are seen as appropriate. For example, in the UK, starting in 2012, the RAE will be replaced by the Research Excellence Framework (REF, 2010). This will consist of a single unified framework for the assessment and funding of research, across all subjects. The new framework will make greater use of quantitative indicators than the RAE, while taking account of key differences between the different disciplines. Similarly, the Australian government decided to abandon the Research Quality Framework and replace it with ERA, which was launched in June 2010. The ERA assessment is conducted through a pure bibliometric approach for the natural and formal sciences[2]. Single research outputs are evaluated by a citation index, relative to world and Australian benchmarks. In the US, there are ranking exercises conducted by the National Research Council, to provide information on the research profile of universities and help them to improve quality through benchmarking. These have also gradually adopted greater use of bibliometric indicators (Hicks, 2009). As a final example, in Italy, the plan is again that the next five-year evaluation exercise VQR will integrate bibliometric analysis with peer review.

Scholars, scientists, policy makers and top managers of research institutions are increasingly involved in debates as to the strong and weak points of these exercises and, in general, of performance based funding (Shattock, 2004; Orr et al., 2007; Strehl et al., 2007). Inquiry has even examined the question of whether incentive schemes can have adverse effects on research (Bhattacharya and Newhouse, 2008; Butler, 2003). An exhaustive analysis of advantages and disadvantages of performance-based approaches to university research funding can be found in Geuna and Martin (2003). While strategic choices should guide funds allocation priorities both at nation and organization levels, once the strategic research areas have been prioritized then the allocation of funds within each area should be based on merit. In fact, in spite of the ongoing debate, there is broad consensus that permanent adoption of performance based funding is desirable, provided its primary goal is to encourage and reward excellence of research in public research organizations (PROs). However, the desired macroeconomic effect could be compromised if internal redistribution of government resources within each PRO does not follow a consistent logic. The desired effects of national evaluation systems for research can result only if a "funds for quality" rule is followed at all levels of PROs' decision-making. However, this merit based re-direction of incoming funds does not necessarily occur. In the UK, the next REF foresees the identification of amounts of funding that are provided as block grants, but universities will then be free to spend the grants as they determine. The next Italian VQR, which like the REF is based on a subset of scientific production from each PRO as a whole, does not provide information on which researchers within the institution contribute most to overall performance. All national assessment exercises that limit the number of research outputs to be submitted by each researcher can at best provide information on the relative quality of researchers based on such limited subset of overall production. It is up to each PRO itself to choose whether to develop internal evaluation systems to identify the most deserving researchers and allocate resources accordingly. However, when we examine the literature, we see that while it abounds with surveys of national systems for performance-based funding, there seem to be few studies of the further extension or effects of such funding systems within the organization and management of PROs. It seems that most PROs likely apply some form of internal performance-based resource allocation, but there is no significant evidence of exhaustive empirical surveys of such systems, while very few operational models have been proposed to inform selective funding allocations to research staff.

---

[2] The peer-review approach is used for the social sciences, arts and humanities.

It seems likely that the lack of contributions on the subject of performance-based resource allocation within PROs is due to the complexity of the potential task. With regard to bibliometric approaches, measurement of performance indicators is greatly affected by availability of data, and by characteristic technical and methodological problems that render robust comparative analyses difficult at the level of individuals. In some countries possible ethical issues associated with individual evaluation could also present a problem.

The objective of this work is to propose a national-scale evaluation support system which could allow individual institutions: i) to identify field strengths and weaknesses, aimed at informing strategic planning; and ii) to assess research performance at individual and departmental levels, in order to optimize funding allocations. The system proposed has so far been used by six Italian universities[3].

The following section presents a brief review of the literature on similar models. The third section describes the methodological details of the model proposed, the dataset used and the indicators taken into consideration. The fourth section provides some elaborations as examples of the application of the methodology, to the Italian case, while the fifth and last section gives a synthesis of the work and the author's comments.

---

[3] University of Rome "Tor Vergata", Milan, Luiss, Pavia, Udine, and Cagliari.

## 2. Large-scale individual research performance evaluation methodologies

This work takes inspiration from the question: is it possible to measure the performance of an individual scientist "A" active in a field "J" and compare it to the one of a scientist "B" active in a field "K"?

Describing the application of journal impact measures in allocating funding among the various faculties at the Delft University of Technology, in the Netherlands, Van Leeuwen and Moed (2002) answer in the affirmative, proposing a measurement system based on the average impact of the scientific production of A and B, standardized with respect to the specificities of J and K, but ignoring the potential difference in productivity between A and B (even though this should represent a fundamental indicator of scientific performance). Previously, Van den Berghe et al., (1998) presented a general methodology applied for a study conducted in the faculties of medicine, science, and pharmaceutical science at three Flemish universities. Rousseau and Smeyers (2000) showed the interesting case of the LUC's research council funding scheme, based on a research evaluation exercise partly grounded on a full-scale scientometric study.

More recently, Costas et al. (2010), after cogently recalling the difficulties and limits of large-scale micro-level research performance analyses, which also refer to our work, propose a general bibliometric methodology for informing the assessment of research performance of individual scientists. They apply their methodology to three research areas of the Spanish National Research Council, totaling 1,064 researchers. The authors set up a bibliometric profile for every researcher, derived from the Web of Science™ (WoS), composed of nine performance variables. Through factor analysis, the nine variables were then reduced to three dimensions: impact, journal quality, and production.

Franceschet (2009) proposes a method to group bibliometric indicators into clusters of highly inter-correlated indicators. Applying his clustering method to the evaluation of a sample of 13 computer science scholars, he clusters 13 indicators into four indexes: i) number of papers, measuring scholar productivity; ii) number of citations, measuring absolute impact of the scholar; iii) average number of citations per paper, measuring relative impact of the scholar; and iv) *m*-quotient[4], measuring enduring impact over time.

The underlying philosophy for the methodology we propose does not involve the clustering approach. Rather than beginning from a large number of indicators (which the proposed elaboration system would be able to measure) and then proceeding to a subsequent clustering or to a final composite indicator as a basis for rankings, the preference is to identify a limited number of indicators that are strongly indicative of the performance dimension for which measurement is desired. It is then left to individual institutions or departments, according to their context of operation, to choose which indicators to actually use and what weight to give each of them. The proposed system has been conceived for large-scale assessments (nation-scale), such as comparing research performance of individual scholars within a field to that of their colleagues in the same nation, in the same or other fields; or of departments of an institution with that of others of the same or other institutions; or of an institution active in a field or discipline with that of other institutions in the country. The development of author-name disambiguated databases of publications in other nations, such as the one underlying our methodology, described below, would offer the useful possibility of international comparisons. The objective of this methodology is to support institutions in the processes of strategic planning, in verifying the effectiveness of policies and initiatives for continuous improvement, in selective funding allocation, etc. Thus to serve these desired ends, the period for assessments must necessarily be brief, on the order of three years.

---

[4] The *m-quotient* is the h-index divided by the research age (Hirsch, 2005).

## 3. The proposed methodology

The proposed methodology applies to the national-scale evaluation of research performance of individual scientists, through measurement of several bibliometric indicators concerning output from research activity. This means that the method considers publications in international journals, but not other forms for codification of the results from scientific activity or other relevant dimensions of university activity, such as teaching and technology transfer. An immediate consequence of this methodology is also that the field of application is limited to the hard sciences, where the use of publications as a proxy for research output gives a high level of representativeness.

The national-scale evaluation of research performance at the level of individual scientists is quite a complex exercise in terms of methodology. It requires an exhaustive census, at the level of individual names, of the scientific production of individual researchers. This presents a formidable task, when using current bibliometric databases such as Elsevier's SCOPUS and Thomson Reuters' Web of Science, in which it is truly difficult to: i) identify and reconcile the varying ways in which authors of publications indicate the name of their "home" institution and ii) fully and properly identify the precise authors of a publication in an automated way. The problem is that in these databases the "authors list" and the "address list" are not fully linked, and as a consequence, whenever the address list indicates two or more institutional affiliations, it is not readily apparent to which one each author belongs. In addition, only the authors' last names and first name initials are reported. When one observes large populations of scientists, the number of homonyms can be very high (in the Italian academic system, we found that 12% of the 60,000 scientists have names that are homonyms) and the disambiguation of names within acceptable margins of error is truly a challenging exercise.

The methodology proposed involves first overcoming the obstacles to identifying authorship, as illustrated in Abramo et al. (2008a) and discussed below. For each scientist in the PRO under observation, it then provides performance ratings for a series of indicators and relative rankings, at a national level, with respect to other colleagues in the same discipline. The rankings, when expressed as percentiles, also permit also comparative analysis of scientists belonging to different fields and disciplines, and, by aggregation, of research groups and departments in the same PRO.

### *3.1 Data sources and field of observation*

The proposed methodology will be applied to the case of Italian universities. The data used in the study are obtained from the Observatory on Public Research in Italy (ORP, 2009), a bibliometric database developed by the authors, which provides a census of international scientific production by PROs in Italy. The ORP is in turn based on the raw data of the National Citation Report of Italy, derived from the Thomson Reuters Web of Science™ (WoS), including conference proceedings. Beginning from these data, and using a complex algorithm[5] for the reconciliation of the authors' affiliations and for the disambiguation of the precise identity of each author, each publication is correctly attributed to the author or authors that wrote it[6].

In Italy, each university researcher must belong to an official scientific disciplinary sector (SDS), and can only belong to one of these SDS. The SDSs in turn compose 14 university disciplinary areas (UDAs). The field of observation for this study consists of the assistant, associate and full professors of Italian universities who belong to the 183 SDSs that compose the "hard sciences". In the Italian case, these correspond to 8 UDAs: Mathematics and computer sciences, Physics, Chemistry, Earth sciences, Biology, Medicine, Agricultural and veterinary sciences, and Industrial and information engineering[7]. These UDAs consist of a total of 34,163 scientists, affiliated with 71 universities. These constitute the dataset for the application of the methodology.

*3.2 Performance indicators*

The basic indicators used to evaluate the performance of individual scientists refer to the quantity and impact of their scientific production. Examining each publication (article or review). recorded in the 2004-2006 period, the evaluation considers the citations it has received up to March 31, 2008. Since the rate of citations is especially sensitive to the discipline involved, we have conducted the analysis by ISI category[8], of which there are 168 for the hard sciences, and defined a standardized quality index for each publication:

*Publication Impact Index* (PII): number of citations (including self-citations[9]) of a publication divided by the average number of citations of all Italian publications[10], of the same type and year, falling in the same ISI category. For instance, a value of 1.40 indicates that the publication was cited 40% more often than the average.

Although the ISI category classification is not perfect, (Leydesdorff, 2008; Bornmann et al. 2008), it provides a clear and consistent definition of fields suitable for automated procedures. After investigating alternative classification methods, Sandström and Sandström (2009) concluded that "there is no simple method e.g., bibliographic coupling, that would be suited for developing a new and better classification," and that "with small fine-tuning, the field definitions and boundaries used by the Thomson Reuters are very well adapted to the needs of a pragmatic evaluative approach".

---

[5] The algorithm is presented in a manuscript which is currently under consideration for publication in another journal. A short abstract is available at
http://www.disp.uniroma2.it/laboratorioRTT/TESTI/Working%20paper/Giuffrida.pdf

[6] At this time, for the identification of authorship of all publications by Italian university researchers indexed in the WoS between 2004 and 2006, the harmonic average of precision and recall (F-measure) is close to 95% (2% sampling error, 98% confidence interval).

[7] "Civil engineering and architecture" is not considered because the WoS does not cover a satisfactory range of research output in this area.

[8] The ISI subject categories are the scientific disciplines that the WoS uses for the classification of publications. The complete list can be seen at http://science.thomsonreuters.com/cgi-bin/jrnlst/jlsubcatg.cgi?PC=D

[9] The authors adhere to the school of thought that a reasonable share of author self-citations is a natural part of scientific communication, and that alarm over author self-citation lacks empirical foundation.

[10] Alternatively, the denominator could be the average number of citations of all WoS indexed publications. In this case the standardization benchmark would be international.

Since the distribution of citations is typically highly skewed in each discipline, we have also used another method for standardization of citations: that of the percentile. The quality index will thus be:

*Publication Impact Ranking* (PIR): ranking of a publication, measured on a 0 – 100 scale, according to the citation distribution of publications of the same type and year falling in the same ISI category. A value of 90 indicates that 90% of the publications of the same year falling in the same ISI category have a lower number of citations than the one under observation.

For the comparative evaluation of performance of individual scientists, the methodology provides a number of indicators that can be measured through the ORP, some of which concern only the quantity produced, others the impact, others the average impact of the scientific production, still others the contribution both to quantity and to impact (a synoptic table is presented in the Annex). We assign the indicators to two categories: the first referring to productivity, the second to average impact.

Productivity indicators[11]
- Productivity (P): total of publications authored by a scientist in the period under observation;
- Fractional Productivity (FP)[12]: total of the contributions to publications authored by a scientist, with "contribution" defined as the reciprocal of the number of co-authors of each publication;
- Scientific Strength, ($SS^{PII}$ or $SS^{PIR}$): the weighted sum of publications authored by the scientist, the weights for each publication being equal to the quality index of the publication (PII or PIR).
- Fractional Scientific Strength, ($FSS^{PII}$ or $FSS^{PIR}$): similar to Fractional Productivity, but referring to Scientific Strength.

More specific elaborations of fractional indicators are given provided for certain disciplines, where the order of the author names has meaning in terms of level of contribution to the publication. For example, in the case of life sciences, the first and last authors are given more weight than the second and the one before last which, in turn, are given more weight than the others.

Average impact indicators
- Quality indexes ($QI^{PII}$ or $QI^{PIR}$): average impact of publications authored by a scientist, i.e. mean values of PII or PIR of publications by a given author.

As can be seen in the following section, each indicator, will be expressed as an absolute or percentile value. The latter serves towards the desired comparison of research performance by scientists that belong to different disciplines.

---

[11] Research productivity by individual scientists is not standardized with respect to effective hours of research nor with respect to other production factors and intangible resources, because of the lack of data that can be attributed to individuals.

[12] More specific indications of fractional productivity could be given for disciplines where the order of the author names conveys a meaning concerning level of contribution to the publication. For example, in the case of Medicine, the first and last authors could be given more weight than the others.

The system would also allow the measurement of the extremely popular h-index, and its variations, but we discourage its use because it is not standardized and because for short assessment periods (3-year window) it is of little use. Moreover, we believe that indicators of impact and volume together bring more information to bear than the single h-index indicator. While each indicator conveys useful information for a decision-maker or individual researcher, we argue that the optimum bibliometric indicator for measure of research performance is that which represents the contribution to advancement of knowledge, i.e. fractional scientific strength. A high value for this indicator can be due, in varying measure, to its determining factors: productivity, average impact of the publications and contribution. Awareness of performance along each of these determinants can be useful for understanding the relative weight of each and for undertaking subsequent intervention for improvement.

## 4. Application

The proposed model of evaluation is based on five simple steps:
a) identification of all university researchers, their home universities and academic rank, for the period under observation;
b) census of the scientific production by each named scientist[13];
c) calculation of bibliometric indicators of productivity and impact, for each scientist;
d) comparison among all scientists of the same SDS and academic rank, and calculation of national percentile of performance (0 being worst, 100 being best) for each indicator;
e) aggregation of performance by research group, department and SDS.

As an example, we present the application of the methodology as a support system for evaluation in the following cases: i) comparison of researchers belonging to the same SDS, within a single university; ii) to different SDSs; iii) comparison of research groups and departments; iv) comparison of the SDSs represented at a university. We refer to publications (articles and reviews) recorded in the 2004-2006 period, and the citations received up to March 31, 2008.

## 4.1 Comparison of scientists within the same SDS

This section presents an example of the comparative evaluation of researchers of a single SDS in one university, in this case the 11 researchers in the BIO/11 SDS (Molecular biology), at the University of Rome "Tor Vergata". Table 1 presents the absolute values for the bibliometric indicators registered for each researcher, while Table 2 presents the relative percentile rankings in comparison with the performance of all Italian university researchers belonging to this SDS[14].

---

[13] The exact authorship of publications could also be subsequently verified by each individual author, to reduce errors and assure the transparency of the evaluation process.
[14] As of December 31, 2005, this SDS had 206 university scientists in all of Italy.

| Scientist ID | P | FP | SS$^{PII}$ | FSS$^{PII}$ | SS$^{PIR}$ | FSS$^{PIR}$ | QI$^{PII}$ | QI$^{PIR}$ |
|---|---|---|---|---|---|---|---|---|
| 1 | 43 | 6.18 | 55.47 | 7.45 | 2859.63 | 384.78 | 1.29 | 66.50 |
| 2 | 15 | 2.75 | 3.86 | 0.78 | 395.04 | 77.60 | 0.26 | 26.34 |
| 3 | 11 | 2.58 | 13.03 | 3.38 | 773.85 | 180.54 | 1.19 | 70.35 |
| 4 | 9 | 1.36 | 5.21 | 0.69 | 468.02 | 65.47 | 0.58 | 52.00 |
| 5 | 9 | 2.57 | 3.51 | 1.02 | 318.55 | 96.45 | 0.39 | 35.40 |
| 6 | 6 | 1.24 | 0.85 | 0.24 | 87.92 | 23.80 | 0.14 | 14.65 |
| 7 | 5 | 0.86 | 1.64 | 0.33 | 158.08 | 31.62 | 0.33 | 31.62 |
| 8 | 4 | 0.41 | 4.57 | 0.40 | 266.73 | 25.18 | 1.14 | 66.68 |
| 9 | 4 | 0.73 | 1.34 | 0.24 | 146.58 | 25.62 | 0.34 | 36.65 |
| 10 | 3 | 0.49 | 0.78 | 0.09 | 103.50 | 12.26 | 0.26 | 34.50 |
| 11 | 1 | 0.14 | 0 | 0 | 0 | 0 | 0 | 0 |

*Table 1: Bibliometric indicators registered for scientists in SDS BIO/11 (Molecular biology) at the University of Rome "Tor Vergata" (2004-2006).*

| Scientist ID | P | FP | SS$^{PII}$ | FSS$^{PII}$ | SS$^{PIR}$ | FSS$^{PIR}$ | QI$^{PII}$ | QI$^{PIR}$ |
|---|---|---|---|---|---|---|---|---|
| 1 | 100 | 100 | 100 | 100 | 100.0 | 99.4 | 87.3 | 72.9 |
| 2 | 94.6 | 93.4 | 59.6 | 67.5 | 71.7 | 79.5 | 22.3 | 19.3 |
| 3 | 87.3 | 92.2 | 90.4 | 95.8 | 89.8 | 97.0 | 84.9 | 79.5 |
| 4 | 80.7 | 78.3 | 70.5 | 65.1 | 77.7 | 70.5 | 51.2 | 51.8 |
| 5 | 80.7 | 91.6 | 57.8 | 75.9 | 64.5 | 83.1 | 32.5 | 25.3 |
| 6 | 69.3 | 74.1 | 27.1 | 41.0 | 30.1 | 42.2 | 15.1 | 9.0 |
| 7 | 63.3 | 59.6 | 40.4 | 45.2 | 41.0 | 50.6 | 27.1 | 22.3 |
| 8 | 53.6 | 32.5 | 66.9 | 51.2 | 59.0 | 43.4 | 84.3 | 74.1 |
| 9 | 53.6 | 53.0 | 36.7 | 39.2 | 39.8 | 44.0 | 28.3 | 27.7 |
| 10 | 44.6 | 38.0 | 25.3 | 24.1 | 32.5 | 26.5 | 22.9 | 24.1 |
| 11 | 20.5 | 15.1 | 0 | 0 | 0 | 0 | 0 | 0 |

*Table 2: National percentile rankings of bibliometric indicators for the scientists in SDS BIO/11 (Molecular biology) at the University of Rome "Tor Vergata" (2004-2006).*

The tables show that there is one researcher who is the national best in the SDS, with 43 publications in the triennium. The second-ranking researcher, with 15 publications, still places in the first decile for productivity, while 9 out of 11 place above the national median for productivity. The lowest ranking scientist of this group registers a single publication, which receives no citations (QI$^{PII}$ and QI$^{PIR}$ both nil). For fractional productivity, four researchers place in the first decile at the national level (ID 1, 2, 3 and 5), while 3 place under the median (ID 8, 10 and 11). The analysis of data for scientific strength does not show any substantial difference to those for productivity.

In reality, since it has been demonstrated that scientific productivity varies with variation in academic rank (Abramo et al., 2008b), the comparison of scientists within a single SDS should actually be conducted at the level of parity in role. In the Italian university system, research personnel are divided in three levels: full, associate and assistant professors. Considering these roles and recalculating the national percentiles according to academic rank, the performance of the 11 researchers in the BIO/11 SDS at "Tor Vergata" presents the situation shown in Table 3. In terms of productivity (P), there is little change in the positioning of the top researchers; however there is a reversal of the positions for researchers with ID 5 and 6, while the full professor with ID 9, who first placed above the national median (53.6), now falls in a much lower percentile (36.2). Also, the assistant professor with ID 8, who first had a national percentile for productivity of 53.6, now achieves a 70.0 ranking. Finally, the associate professor with ID 3 now tops the national rankings for fractional scientific strength (FSS$^{PII}$ and FSS$^{PIR}$) while previously this researcher's national percentile rankings for these two indicators were respectively 95.8 and 97.

| Scientist ID | Acad. rank | P | FP | $SS^{PII}$ | $FSS^{PII}$ | $SS^{PIR}$ | $FSS^{PIR}$ | $QI^{PII}$ | $QI^{PIR}$ |
|---|---|---|---|---|---|---|---|---|---|
| 1 | Full | 100 | 100 | 100 | 100 | 100 | 98.3 | 81.0 | 70.7 |
| 2 | Full | 87.9 | 84.5 | 43.1 | 50.0 | 53.4 | 60.3 | 13.8 | 13.8 |
| 3 | Associate | 91.3 | 97.8 | 95.7 | 100 | 95.7 | 100 | 93.5 | 84.8 |
| 4 | Associate | 87.0 | 87.0 | 78.3 | 76.1 | 84.8 | 80.4 | 54.3 | 58.7 |
| 5 | Full | 63.8 | 82.8 | 39.7 | 60.3 | 44.8 | 69.0 | 24.1 | 15.5 |
| 6 | Associate | 73.9 | 78.3 | 23.9 | 43.5 | 28.3 | 43.5 | 10.9 | 8.7 |
| 7 | Associate | 65.2 | 63.0 | 45.7 | 52.2 | 45.7 | 52.2 | 28.3 | 26.1 |
| 8 | Assistant | 70.0 | 46.7 | 75.0 | 63.3 | 71.7 | 56.7 | 85.0 | 75.0 |
| 9 | Full | 36.2 | 32.8 | 22.4 | 25.9 | 22.4 | 29.3 | 20.7 | 20.7 |
| 10 | Assistant | 61.7 | 55.0 | 35.0 | 31.7 | 46.7 | 38.3 | 30.0 | 30.0 |
| 11 | Assistant | 31.7 | 23.3 | 0 | 0 | 0 | 0 | 0 | 0 |

*Table 3: National percentile rankings of scientists in SDS BIO/11 (Molecular biology) at the University of Rome "Tor Vergata", considering their academic rank (2004-2006).*

### 4.2 Comparison of scientists from different SDSs

The use of national percentiles also permits comparisons in performance between scientists belonging to different SDSs. Table 4 presents the case of two scientists in the Physics area at the University of Milan.

| | Scientist A FIS/03 | | Scientist B FIS/06 | |
|---|---|---|---|---|
| Index | *Abs. value* | rank% | *Abs. value* | rank% |
| P | 15 | 74,6 | 12 | 100 |
| FP | 3.97 | 76.6 | 2.08 | 89.1 |
| $SS^{PII}$ | 12.52 | 75.1 | 11.92 | 100 |
| $FSS^{PII}$ | 3.12 | 80.4 | 1.75 | 93.5 |
| $SS^{PIR}$ | 923.62 | 78.9 | 618.25 | 95.7 |
| $FSS^{PIR}$ | 241.70 | 83.7 | 98.99 | 91.3 |
| $QI^{PII}$ | 0.84 | 71.8 | 0.99 | 95.7 |
| $QI^{PIR}$ | 61.58 | 76.3 | 51.52 | 82.6 |

*Table 4: Comparison of bibliometric performance of two scientists at the University of Milan.*

Scientist A, assistant professor in FIS/03 (Physics of matter), produced 15 publications in the triennium under observation. In comparison with 145 colleagues in the same SDS and the same academic rank, a quarter of these showed greater productivity. Meanwhile, Scientist B was assistant professor in FIS/06 (Earth physics and atmospheric environment). His 12 publications over the triennium place him at the top of national rankings for productivity. For fractional productivity, the national percentile ranking for Scientist A (76.6) is again lower than that for Scientist B (89.1), in spite of the fact that the absolute value for performance by Scientist A is greater than that for Scientist B. The same situation occurs for scientific strength: for example, the absolute value for $SS^{PIR}$ achieved by Scientist A (923.62) is higher than that of Scientist B (618.25), but the ranking of national percentiles is reversed: 78.9 for Scientist A compared to 95.7 for Scientist B. It is clear that the simple comparisons of absolute values of indicators can lead to erroneous conclusions concerning the relative performance of these two scientists.

However, the use of national percentile rankings calculated with respect to the distributions within the SDS to which they belong permits a robust comparison between scientists operating in disciplines that are very unlike in terms of "fertility" of publication and patterns of citation.

**4.3 Comparisons among research groups and departments**

The example presented in Section 3.2 shows how comparisons of performance can be made among researchers belonging to different SDSs. Using simple aggregation of standardized bibliometric measures it is thus readily possible to proceed to comparisons of heterogeneous research groups. This method provides universities with a highly flexible evaluation framework, and thus permits them to formulate incentive systems based on the performance of individual scientists, research groups, or formal organizational units, such as departments. We will first refer to the case of research groups, which are often informal aggregations of a small number of scientists who share an interest in a specific line of scientific investigation. As an example we will refer to the cases of two research groups in a single university, both in the area of Physics. The first, composed of 6 scientists belonging to 3 different SDSs, carries out research in the general field of optics and spectroscopy. The second, composed of 8 scientists belonging to 5 SDSs, focuses on high energy physics. Table 5 presents the national percentile rankings of bibliometric indicators for each scientist belonging to these two groups. Again, the percentile rankings for each indicator are calculated in comparison to the performance of all Italian university scientists belonging to the same SDS and with the same academic ranking. Next, Table 6 presents the mean values of the national percentile rankings for the members of each group: whatever indicator is examined, it can be seen that the performance of Group 2 is always superior to that of Group 1.

| SDS code* / Indicator | Group 1 - *Optics and spectroscopy* | | | | | | Group 2 - *High energy physics* | | | | | | | |
|---|---|---|---|---|---|---|---|---|---|---|---|---|---|---|
| | FIS/01 | FIS/01 | FIS/03 | FIS/03 | FIS/03 | INF/01 | FIS/01 | FIS/01 | FIS/01 | FIS/04 | FIS/04 | ING-IND/33 | ING-INF/02 | FIS/03 |
| P | 43.7 | 60.1 | 60.3 | 64.1 | 64.1 | 33.3 | 91.3 | 60.1 | 29.5 | 67.9 | 32.1 | 90.7 | 37.2 | 77.4 |
| FP | 33.8 | 62.8 | 55.5 | 47.3 | 73.0 | 26.4 | 89.0 | 55.5 | 19.9 | 53.2 | 24.4 | 85.0 | 19.3 | 64.6 |
| $SS^{PII}$ | 44.2 | 41.1 | 77.4 | 54.5 | 44.5 | 44.8 | 91.3 | 65.8 | 36.4 | 96.8 | 25.6 | 75.7 | 37.2 | 91.1 |
| $FSS^{PII}$ | 43.7 | 48.2 | 73.0 | 45.3 | 57.0 | 35.7 | 83.7 | 64.4 | 29.4 | 92.3 | 25.6 | 71.0 | 16.6 | 84.5 |
| $SS^{PIR}$ | 44.5 | 52.4 | 67.4 | 60.6 | 52.7 | 43.5 | 81.9 | 70.4 | 33.5 | 78.8 | 26.3 | 78.5 | 31.7 | 84.0 |
| $FSS^{PIR}$ | 42.2 | 57.9 | 65.1 | 46.6 | 67.2 | 39.8 | 78.8 | 71.7 | 28.4 | 72.4 | 25.0 | 71.0 | 13.1 | 72.5 |
| $QI^{PII}$ | 58.6 | 33.4 | 87.0 | 48.6 | 32.1 | 56.7 | 79.5 | 75.0 | 74.2 | 100.0 | 37.8 | 63.6 | 50.3 | 93.9 |
| $QI^{PIR}$ | 56.0 | 42.0 | 84.7 | 48.1 | 32.1 | 56.9 | 43.1 | 91.6 | 89.8 | 98.1 | 30.8 | 63.6 | 26.2 | 89.6 |

*Table 5: National percentile rankings for the scientists of two Physics research groups at an Italian University (2004-2006).*
* *FIS/01 = Experimental physics; FIS/03 = Physics of matter; FIS/04 = Nuclear and subnuclear physics; INF/01 = Computer science; ING-IND/33 = Electrical systems for energy; ING-INF/02 = Electromagnetic fields.*

| Indicator | Group 1 - *Optics and spectroscopy* | Group 2 - *High energy Physics* |
|---|---|---|
| P | 54.3 | 60.8 |
| FP | 49.8 | 51.4 |
| $SS^{PII}$ | 51.1 | 65.0 |
| $FSS^{PII}$ | 50.5 | 58.4 |
| $SS^{PIR}$ | 53.5 | 60.6 |
| $FSS^{PIR}$ | 53.1 | 54.1 |
| $QI^{PII}$ | 52.7 | 71.8 |
| $QI^{PIR}$ | 53.3 | 66.6 |

*Table 6: Average of national percentiles for the scientists of two physics research group at an Italian university (2004-2006).*

The same procedure can be applied at the departmental level, which is the formal organizational unit to which a PRO typically assigns research activity. Table 7 presents the example of the membership, by SDS, for the research staff of the Department of Physics, University of Milan: the department includes 97 researchers, of which 88 belong to 8 SDSs of the Physics UDA, 6 to the Industrial and information engineering UDA, and the remaining 3 to SDSs of the Mathematics and computer science, Chemistry, and Medicine UDAs. The SDSs these researchers belong to are extremely variable in terms of publication fertility (column 4): the mean value of papers per author per year for each SDS in the period under consideration ranges from a minimum of 0.01 to 2.78. This variation does not present an obstacle if, once again, the evaluation proceeds by comparison between percentile rankings of performance for each researcher with respect to national colleagues in the same SDS and with the same academic rank. Table 8 presents the values of such rankings (for the top ten scientists in terms of productivity) in the department under consideration.

With this level of analysis, the head of a university department thus has access to ratings and rankings that, among others, can potentially support decisions concerning assignment of funding among department members.

At a higher level, the aggregations of percentile rankings for each researcher in a department permit the arrival at values of performance that can be used in comparing departments at a university, and thus in funding decisions taken by the administration of a faculty containing a number of departments. Table 9 presents the case of two departments in an Italian university.

| SDS code | SDS name | Research staff | Publication intensity |
|---|---|---|---|
| FIS/03 | Physics of matter | 21 (21.6%) | 2.78 |
| FIS/01 | Experimental physics | 18 (18.6%) | 1.56 |
| FIS/02 | Theoretical physics | 17 (17.5%) | 1.77 |
| FIS/04 | Nuclear and subnuclear physics | 14 (14.4%) | 1.54 |
| FIS/07 | Applied physics | 7 (7.2%) | 1.41 |
| FIS/05 | Astronomy and astrophysics | 6 (6.2%) | 2.45 |
| ING-INF/01 | Electronics | 6 (6.2%) | 1.47 |
| FIS/08 | History of physics | 3 (3.1%) | 0.37 |
| FIS/06 | Earth physics and atmospheric environment | 2 (2.1%) | 0.98 |
| CHIM/03 | General and inorganic chemistry | 1 (1.0%) | 2.04 |
| INF/01 | Computer science | 1 (1.0%) | 1.01 |
| M-PED/01 | General and social pedagogy | 1 (1.0%) | 0.01 |
| | Total | 97 | |

*Table 7: Research staff of the Department of Physics, University of Milan (2004-2006).*

| Scientist ID | SDS | P | FP | SS$^{PII}$ | FSS$^{PII}$ | SS$^{PIR}$ | FSS$^{PIR}$ | QI$^{PII}$ | QI$^{PIR}$ |
|---|---|---|---|---|---|---|---|---|---|
| 1 | FIS/01 | 100 | 61.6 | 100 | 70.5 | 100 | 62.9 | 86.3 | 83.9 |
| 2 | FIS/04 | 99.5 | 80.5 | 98.9 | 73.5 | 99.5 | 73.5 | 66.5 | 66.5 |
| 3 | FIS/01 | 99.2 | 82.8 | 98.4 | 78.0 | 98.9 | 77.8 | 76.5 | 69.3 |
| 4 | FIS/04 | 98.9 | 100 | 92.4 | 93.5 | 97.3 | 97.8 | 61.6 | 51.4 |
| 5 | FIS/04 | 98.4 | 67.6 | 94.1 | 65.4 | 97.8 | 69.7 | 68.6 | 64.3 |
| 6 | FIS/06 | 96.9 | 90.6 | 100 | 95.3 | 98.4 | 93.8 | 96.9 | 90.6 |
| 7 | FIS/04 | 96.8 | 61.1 | 83.2 | 63.2 | 95.1 | 67.0 | 51.4 | 60.5 |
| 8 | FIS/04 | 96.8 | 74.1 | 89.7 | 67.6 | 91.9 | 68.6 | 69.7 | 50.3 |
| 9 | FIS/03 | 96.5 | 83.3 | 96.3 | 91.0 | 98.0 | 91.6 | 82.1 | 90.4 |
| 10 | FIS/07 | 95.0 | 96.4 | 81.4 | 82.5 | 80.3 | 84.2 | 55.1 | 44.0 |
| … | | | | | | | | | |

*Table 8: National percentile rankings of performance indicators for the top 10 scientists (for productivity) in the Department of Physics, University of Milan (2004-2006).*

In the Inorganic Chemistry department there are 34 researchers belonging to only 2 SDSs. In the Pharmacology department there are 47 researchers belonging to 5 different SDSs. The average performance of the researchers in Pharmacology is invariably higher than that of those in the other department. For example, in terms of productivity (P) the average national percentile for the researchers in Inorganic Chemistry is 42, while for those in Pharmacology the average percentile is 74.2. For the dimensions of fractional productivity and qualitative impact of publications, the researchers in Pharmacology again achieve a higher average ranking than those in Inorganic Chemistry (71.1 versus 39.9 for FP; 73.1 versus 40.8 for FSS$^{PII}$, etc.). This example again highlights the importance of carrying out comparisons among scientists that belong to the same SDS, and also to the same academic rank, to eliminate potential distortions linked to the varying compositions of the personnel in each department being evaluated.

| Department | Inorganic chemistry | Pharmacology |
|---|---|---|
| Research staff | 34 | 47 |
| Number of SDSs | 2 | 5 |
| P | 42.0 | 74.2 |
| FP | 39.9 | 71.1 |
| SS$^{PII}$ | 40.9 | 74.3 |
| FSS$^{PII}$ | 40.8 | 73.1 |
| SS$^{PIR}$ | 40.5 | 73.8 |
| FSS$^{PIR}$ | 39.9 | 72.8 |
| QI$^{PII}$ | 44.9 | 68.6 |
| QI$^{PIR}$ | 46.2 | 65.1 |

*Table 9: Average of national percentiles for the scientists of two departments at the University of Milan.*

With further aggregation of performance measures, it would be possible to arrive at comparison of larger administrative units, such as entire colleges or schools within the same university.

## 4.4 Evaluation of SDSs

This section of the paper provides a final example of the application of the proposed methodology to the case of comparing the SDSs within a single university. This application is particularly interesting for strategies of recruitment, considering that in such situations it would be very useful for a university to know the status of its various disciplines (SDSs). For example, if a university were to result as weak in a particular SDS that is considered strategic, it could find it more useful to insert a respected senior scientist, able to strengthen the SDS, rather than a junior scientist. The situation could be the contrary for a strong SDS: a junior scientist could quickly grow and benefit from the accumulated knowledge and guidance offered by seniors within a strong SDS. It can thus be very useful to know the positioning of a university as concerns its various disciplines of activity (SDSs). The analysis by SDS involves some methodological differences compared to the previous applications. Because the individual SDSs are intrinsically homogenous, the preliminary step of the analysis is the simple aggregation of the scientific production of the researchers that compose them. The indicators of productivity can thus be calculated on the basis of this "portfolio", dividing the overall output by the number of researchers that compose the SDS. The impact indicators can also be calculated through the simple ratio between Scientific Strength and the number of publications in the SDS. As an example of this methodology, Table 10 presents the evaluation of the 6 SDSs of a small university, the International School for Advanced Studies of Trieste.

| SDS code* | BIO/09 | MAT/05 | FIS/05 | FIS/03 | MAT/07 | FIS/02 |
|---|---|---|---|---|---|---|
| Research staff | 5 | 10 | 8 | 11 | 5 | 9 |
| Papers | 68 | 67 | 130 | 189 | 22 | 49 |
| P | 100 | 98.0 | 94.1 | 93.9 | 55.3 | 20.0 |
| FP | 100 | 95.9 | 100 | 63.6 | 47.4 | 26.7 |
| $SS^{PII}$ (per scientist) | 100 | 100 | 94.1 | 97.0 | 71.1 | 20.0 |
| $FSS^{PII}$ (per scientist) | 100 | 100 | 100 | 84.8 | 65.8 | 30.0 |
| $SS^{PIR}$ (per scientist) | 100 | 98.0 | 94.1 | 97.0 | 63.2 | 30.0 |
| $FSS^{PIR}$ (per scientist) | 100 | 100 | 100 | 69.7 | 52.6 | 33.3 |
| $QI^{PII}$ (per paper) | 88.1 | 100 | 88.2 | 93.9 | 68.4 | 46.7 |
| $QI^{PIR}$ (per paper) | 85.7 | 98.0 | 100 | 97.0 | 65.8 | 66.7 |

*Table 10: National percentiles for the SDSs of the International School for Advanced Studies, Trieste (2004-2006).*
*\* BIO/09 = Physiology; MAT/05 = Mathematical analysis; FIS/05 = Astronomy and astrophysics; FIS/03 = Nuclear and subnuclear physics; MAT/07 = Mathematical physics; FIS/02 = Theoretical physics.*

The excellent performance of this university's BIO/09 SDS (Physiology) is readily apparent, with its leadership in national rankings for all of the 8 indicators of productivity. The performance of the SDS MAT/05 (Mathematical analysis) is also excellent, above the 95th percentile in the sector for all measures of performance. FIS/05 (Astronomy and astrophysics) and FIS/03 (Nuclear and sub-nuclear physics) also register excellent bibliometric performances for all indicators, with the possible exception of the contribution indicators for FIS/03 (FP, $FSS^{PII}$ and $FSS^{PIR}$), which is a sign that this SDS at this university tends to collaborate more than others with external research organizations, compared to the national mean for the SDS. The last two SDSs seem to achieve a lesser performance. The SDS for FIS/02 (Theoretical physics) actually places in the last national quintile in terms of publications per scientist (P), and under the national median for all other indicators (except $QI^{PIR}$). Such strengths and weaknesses analysis at the sectorial level, as seen here, could help to inform strategic planning, strategic control, recruitment choices, etc.

Since the methodology permits comparative performance evaluation of all the researchers of a nation, it is possible to extrapolate the top scientists (for example the top 10%) for each indicator and then, for every university or SDS, to measure the concentration of top scientists. It is also possible to assess research groups, departments, SDSs, etc. by number of publications with standardized impact above a certain threshold. For example, let us consider the top 1% most cited publications in each ISI category. Table 10 presents the assessment of the 10 SDSs falling in the UDA Biology of the University of Milan. In this case, the performances are measured on the basis of the top 1% most cited publications only. In terms of productivity the SDS BIO/06 ranks first: the 16 scientists belonging to it produced 3 such publications in the period under observation (P=0.063). BIO/14 (P=0.057), ranks second while BIO/10, ranks last (P = 0.005), with only 1 most cited paper produced by its research staff. It can be noted that a number of rankings are correlated, given the specific subset on which they are based.

| SDS | BIO/01 | BIO/04 | BIO/06 | BIO/07 | BIO/09 | BIO/10 | BIO/11 | BIO/12 | BIO/14 | BIO/17 |
|---|---|---|---|---|---|---|---|---|---|---|
| Research staff | 8 | 11 | 16 | 8 | 56 | 62 | 13 | 10 | 76 | 8 |
| Top papers | 1 | 1 | 3 | 1 | 2 | 1 | 2 | 1 | 13 | 1 |
| P | 0,043 | 0,031 | 0,063 | 0,043 | 0,012 | 0,005 | 0,051 | 0,032 | 0,057 | 0,043 |
| FP | 0,004 | 0,005 | 0,008 | 0,009 | 0,004 | 0,000 | 0,011 | 0,003 | 0,010 | 0,002 |
| $SS^{PII}$ (per scientist) | 0,501 | 0,069 | 0,742 | 0,449 | 0,048 | 0,039 | 0,451 | 1,058 | 0,595 | 0,596 |
| $FSS^{PII}$ (per scientist) | 0,042 | 0,012 | 0,113 | 0,090 | 0,011 | 0,001 | 0,136 | 0,088 | 0,093 | 0,030 |
| $SS^{PIR}$ (per scientist) | 4,309 | 3,114 | 6,220 | 4,348 | 1,183 | 0,535 | 5,102 | 3,226 | 5,649 | 4,313 |
| $FSS^{PIR}$ (per scientist) | 0,359 | 0,519 | 0,834 | 0,870 | 0,370 | 0,018 | 1,063 | 0,269 | 0,986 | 0,216 |

*Table 10: Bibliometric indicators registered for SDS of the Biology UDA at the University Milan based only on top 1% cited papers (2004-2006).*

* *BIO/01 = General Botanics; BIO/04 = Vegetal Physiology; BIO/06 = Comparative Anatomy and Citology; BIO/07 = Ecology; BIO/09 = Physiology; BIO/10 = Biochemistry; BIO/11 = Molecular Biology; BIO/12 = Clinical Biochemistry and Biology; BIO/14 = Pharmacology; BIO/17 = Histology*

## 5. Conclusions

The literature abounds with surveys of national systems for performance-based funding of higher education institutions, while there seem to be very few analyses of the effects of such funding systems on organizational and managerial arrangement of PROs. However, this latter subject of inquiry is important, since the desired macroeconomic effects of a national centralized allocation system can be compromised if internal redistribution within each PRO does not follow a similar logic, under which institutional "revenues" are re-directed to "revenue generators". Yet PROs have objective difficulty in comparing the scientific performance of scientists who publish in different disciplines, which are characterized by different intensities of publication and rates of citation. This work proposed a decision support system based on large-scale measurement of the bibliometric performance of individual researchers, offered as an aid for resource allocation and strategic planning in public research organizations. The system is based on the comparison of production of over 30,000 individual scientists, after standardizing for the intensity of citation in the fields of publication. This methodology overcomes the traditional limits of bibliometric analyses and permits robust rankings at the level of individual scientists, fields and departments, which can then be very heterogeneous in terms of field of research. In comparison to other models of assessment the proposed methodology offers for the hard sciences a series of advantages:

- objectivity, rapidity and low cost of implementation when compared to the classic peer review approaches that some universities adopt for internal evaluations, which also present other relative weaknesses, including difficulty in pushing their application to the level of single individuals;
- an exhaustive field of observation, permitting ready and efficient application to the hard sciences, but also relevant to some sectors of social sciences in which bibliometrics is appropriate;
- sophistication of the indicators applied, considering that the literature up until now has featured numerous analyses based on mean impact, while large-scale analyses based on productivity have been almost inexistent;
- robustness of the rankings obtained, considering that the measurements take account of differences in intensity of publication and citation among the various sectors considered, and also of the level of employment or academic ranking of the scientists.

The wide range of performance indicators considered in the model permit the user institutions to assign appropriate weights, in function of the disciplines being considered and the strategic aims of the institution. PROs can also integrate other information with the proposed indicators, such as information on output (patents, databases, agreements, etc.) and inputs (resources), in order to further refine their comparative evaluations. It is also clear that the proposed system offers flexibility in application in support of various decision-making processes (especially in funding and recruiting), and at various organizational levels.

Differently from the Costas et al. (2010) methodology[15], we use a lower number of indicators, in particular we exclude the h-index, and types of research output, in particular we do not consider patents, but we carry out comparisons of performance at a larger scale (34,163 researchers) and at a more micro level of analysis (183 disciplines). The two methodologies probably reflect both slightly different philosophies of evaluation and availability of instruments to carry out comparative performance assessment. The former emphasizes the richness of indicators and types of research output; the latter the amplitude of the benchmark for comparative assessment and the limitation of distortions due to the different citation intensity of disciplines.

---

[15] The authors note that the work by Costas et al. did not inspire the current work, since it came to their awareness only at the moment that the current paper was submitted for publication.

**ANNEX: Indicators of research performance at individual level**

| Cat. | Acr. | Title | Definition |
|---|---|---|---|
| Productivity | P | Productivity | Total of publications authored by a scientist in the period under observation |
| Productivity | FP | Fractional Productivity | Total of the contributions to publications authored by a scientist, with "contribution" defined as the reciprocal of the number of co-authors of each publication |
| Productivity | $SS^{PII}$ | Scientific Strength (based on PII*) | Weighted sum of publications authored by the scientist, the weights for each publication being equal to the PII |
| Productivity | $FSS^{PII}$ | Fractional Scientific Strength (based on PII) | Similar to Fractional Productivity, but referring to Scientific Strength |
| Productivity | $SS^{PIR}$ | Scientific Strength (based on PIR**) | Weighted sum of publications authored by the scientist, the weights for each publication being equal to the PIR |
| Productivity | $FSS^{PIR}$ | Fractional Scientific Strength (based on PIR) | Similar to Fractional Productivity, but referring to Scientific Strength |
| Average impact | $QI^{PII}$ | Quality index (based on PII) | Average impact of publications authored by a scientist, i.e. mean values of PII of publications by a given author |
| Average impact | $QI^{PIR}$ | Quality index (based on PIR) | Average impact of publications authored by a scientist, i.e. mean values of PIR of publications by a given author |

*\* Publication impact index: number of citations (including self-citations) of a publication divided by the average number of citations of all Italian publications, of the same type and year, falling in the same ISI category.*

*\*\* Publication impact ranking: ranking of a publication, measured on a 0 – 100 scale, according to the citation distribution of publications of the same type and year, falling in the same ISI categor*